\documentclass{optica-article}

\journal{opticajournal}

\articletype{Research Article}

\usepackage{lineno}

\begin{document}

\title{Design of a fast speckle wavemeter with optical processing}

\author{Lucas R. Mendicino,\authormark{1,2} and Christian T. Schmiegelow\authormark{1,2,*}}

\address{\authormark{1} Universidad de Buenos Aires, Facultad de Ciencias Exactas y Naturales, Departamento de Física. Buenos Aires, Argentina.
\\
\authormark{2} CONICET - Universidad de Buenos Aires, Instituto de Física de Buenos Aires (IFIBA). Buenos Aires, Argentina}

\email{\authormark{*}schmiegelow@df.uba.ar} 



\begin{abstract}
    We present a design concept for a speckle based wavemeter that combines high spectral resolution and fast response times. Our device uses a fixed disperse medium with small coherence length as an optical pre-processor and a series of programmable optical elements as post-processor.
    The pre-processor generates a complex speckle pattern with a given correlation length, then the post-processor transforms the optical field to a simple binary pattern with intensity proportional to the wavelength deviation from a reference value. We show how to construct a device which can be trained to produce an electrical signal on a balanced detector proportional to the wavelength.
    Also, we demonstrate that the device can operate from sub-picometer up to nanometer resolution using a pre-processor with correlation length in the picometer range. 
    More generally, our results show how the use of a programmable optical post-processor with low spectral resolution can be enhanced by a fixed pre-procesor with a higher one.
\end{abstract}

\section{Introduction}

Modern wavelength-meters, working on the principle of interference, can be divided into two categories: those which diffract in a random fashion (speckle based) and those that use ordered structures (interference pattern based). In both cases, the diffraction pattern is usually read by an array of detectors which is then digitized and processed to determine the wavelength. This stage sets a limit to the readout speed which is usually well above the millisecond on commercially available wavelength-meters, and even using a fast-reading camera the speckle based wavemeter have achieved frequencies for laser stabilization that are below 50~kHz~\cite{metzger2017harnessing}. Here, we propose replacing the complex detection and digital post-processing with an optical post-processor and a balanced binary detector which easily operate below microsecond timescales. This is of particular interest for laser-frequency stabilization where such a wavelength-meter could provide not only absolute frequency determination but also stabilization timescales in the MHz regime and resolutions well below the picometer, comparable to Fabry-Perot cavities. Moreover our idea could be immediately extended to generate programmable wavelength division multiplexers.

Speckle based wavemeters have exploded several ways to generate optical randomness~\cite{cao2017perspective, wan2021review}. To mention a few, multimode fiber~\cite{redding2012using} and integrating spheres~\cite{metzger2017harnessing} that showed correlation lengths in the order of a few~pm. Thinking on commercial applications some on-chip developments have shown great results in designs such as evanescently coupled multimode fibers~\cite{redding2016evanescently} and tailored dissordered photonic chips~\cite{redding2013compact,hartmann2020broadband}. On more recent advances, an integrated multimode waveguide connected to a photonic lantern~\cite{yi2020integrated} and even microring resonators coupled to interferometers of nanometer scale~\cite{zhang2021compact} have been used as compact and integrated wavelength-meters.
These techniques have achieved resolutions from a few picometers to a few attometers with scaling complexity of the system. However, as the complexity scales the time required to process the images grows, this represents a constraint in the development of speckle based stabilization systems.

To infer the wavelength of light from a speckle pattern, several linear algebra analysis techniques have been used, such as transmission matrix~\cite{cao2017perspective}, principal component analysis~\cite{metzger2017harnessing}, Poincaré descriptors~\cite{o2020high} and even in the last years deep learning techniques have been used with the same purpose~\cite{gupta2020deep}.
These techniques require a calibration stage that associates several speckle patterns with several known wavelengths, this information is digitally stored. Following, when a real measurement is carried out, the patterns generated are compared with the ones stored and the algorithm gives as an output the wavelength associated to the pattern. These steps require processing time which can be sped up drastically with an optical processor. 

In this work we propose the replacement of the digital post-processor by an optical one. The optical post-processor is composed by a set of programmable phase plates which are trained to convert the complex wavelength-dependent speckle patterns into a binary outcome proportional to the wavelength deviation $\Delta\lambda^\mathit{tr}$ from a chosen wavelength $\lambda_0$. This signal can then be read on a balanced photodetector providing an electrical readout of the wavelength with a time constant only limited by the bandwidth of a single or balanced detector.

To train the programmable planes we use a wavefront matching technique~\cite{sakamaki2007new}. This technique has shown  the capability to efficiently convert many different input beams into a variety of chosen output beams. For example, it was shown that N different Laguerre-Gaussian beams could be sorted into spatially separated output beams, all with a Gaussian profile~\cite{fontaine2019laguerre}. More recent results on the decomposition of light using spatial light modulators include the resolution of the spatial, spectral, temporal and polarisation state of the outcome light of a VCSEL (Vertical-cavity surface-emitting laser)~\cite{ploschner2022spatial}, the design and fabrication of an ultra-broadband optical hybrid using a modification of the wavefront matching technique~\cite{zhang2020ultra} and the design of a hybrid mode and wavelength division multiplexing device~\cite{wei2022parallel}.

Programmable spatial light modulators can also be trained as optical processors. Moreover, the use of a randomising media, with higher dimension than the programmable planes can provide higher control and versatility on the overall optical processor as well as better resilience to fabrication imperfections. In particular, it was shown, that the use of a multi mode fiber, in combination with a set of programmable planes can be used to implement arbitrary gates on photonic modes with dimension up to 7~\cite{marcucci2020programming}.

In this work, we show in section~\ref{sec: pre} that a wavefront matching technique can be used to train a set of spatial light modulators to distinguish beams with the same spatial structure but different frequency and how the inclusion of a fixed randomizing media as a pre-processor enhances the resolution of the programmable post-processor. In section~\ref{sec: algorithms} we show how to use these ideas to construct a full wavemeter with sub-picometer resolution and wide spectral response.

\section{Methods}\label{sec: pre}

\begin{figure}[ht]
    \centering
    \includegraphics[width=0.7\columnwidth]{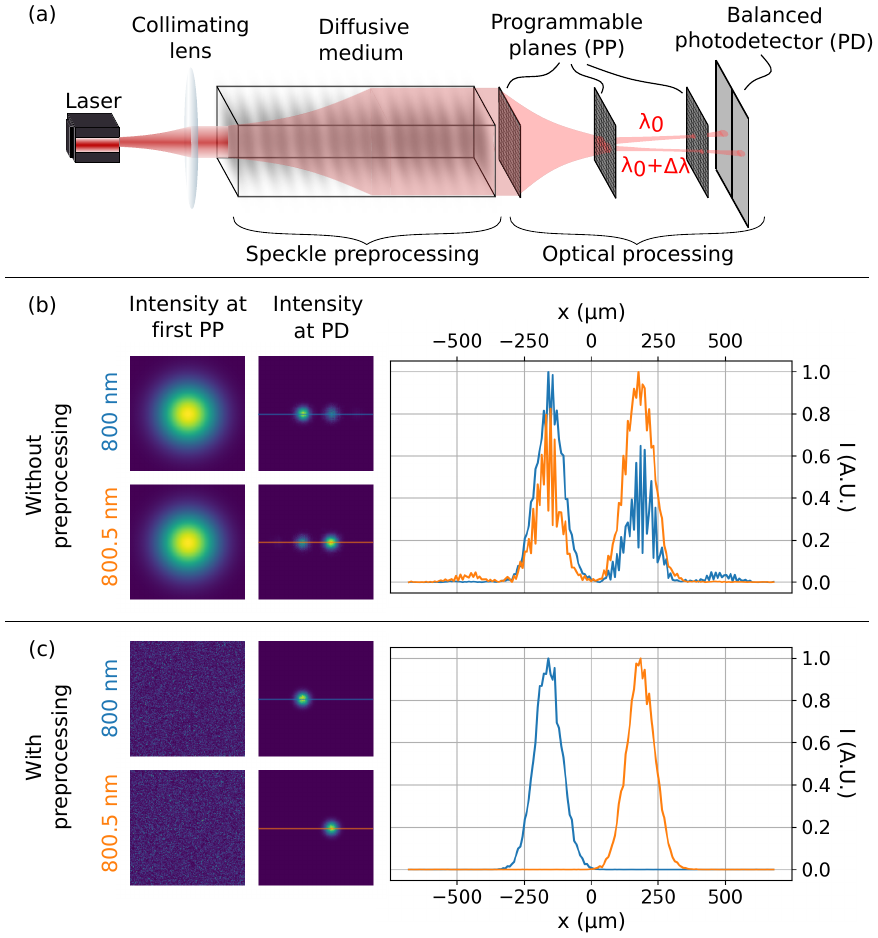}
    \caption{General overview of the method. (a) Scheme of the proposed setup. A laser beam is collimated and then propagates through a diffusive medium in order to generate a speckle pattern as a consequence of the auto-interference of the light generated by the randomization of the wavefront. Then the light propagates through three programmable planes, previously trained to send light with wavelength $\lambda_0$ to one side of the PD and light with wavelength $\lambda_0 + \Delta \lambda$ to the other one. (b, c) Demonstration of the method in the case that $\Delta \lambda =$~$0.5$~nm. In (b) the light is not pre-processed because of the difussive medium is not present, so the planes are trained with a Gaussian beam and in (c) the light is pre-processed so the light that gets to the first programmable plane shows a speckle pattern. The first column is the intensity of light that arrives at the first of the programmable planes, the second one is the intensity at the plane of the photodetector~(PD), i.e. the outcome. The plots in the right hand show the intensity obtained through the line drawn in the PD plots. Blue and orange lines represent 800~nm and 800.5~nm respectively. It can be seen that the use of a speckle generator as a light pre-processor allows us to identify the wavelength of the incoming light with a cleaner signal at the PD.}
    \label{fig:Method}
\end{figure}

The proposed scheme for the realization of the wavemeter can be seen in fig.~\ref{fig:Method}(a). It has two main and independent sections: the speckle pre-processor and the optical post-processor. To test the performance of the device we simulated its behaviour using a coherent beam propagation method implemented in the library Light~Pipes~\cite{lightpipes} for Python.

In the pre-processing stage we generate a speckle pattern. The pre-processor is modeled by a square waveguide that has equidistant diffusive planes. The objective of the diffusive planes is to randomize the wavefront of the light as it passes trough every plane. In this way it is possible to generate speckle patterns with different correlation length by adjusting the characteristic distances between the planes, the number of planes and the distance between them. The physical realization of such a pre-processor could be done with any of the experimental methods mentioned above, given that the output light is a speckle pattern arising from the random auto-interference with some characteristic coherence length. Here, we use $N$~=~50 planes of section $4.096$~mm~$\times$~$4.096$~mm, separated by $50$~cm, with a characteristic random scattering phase element of size 512~$\mu$m~$\times$~512~$\mu$m. Also, we consider lossless conditions by assuming that light which reaches the borders of the simulation is reflected inwards. This way, by changing parameters such as the distance between the planes, the amount of planes and the size of the random scattering phase element we can achieve speckle patterns with correlation lengths from hundreds of~fm to hundreds of~pm.

At the optical post-processing stage, we prepared three programmable planes, whose pixel area were chosen to be $\sim60~\mathrm{\mu m}^2$, similar to typical commercial SLMs; the phase of every pixel can be programmed by setting an external electric potential. Thinking on a possible experimental implementation, we tried to use as few planes as possible. We ended up using 3 planes because for 1 plane the back and forth propagation of the wavefront matching algorithm (see section~\ref{sec: PPT}) was not possible and for 2 planes the change in the propagation in each one was too abrupt and the algorithm was not always able to converge to the expected wavefront. The distance between the programmable planes was 9~cm and their size was $4.096~\mathrm{mm} \times 4.096~\mathrm{mm}$. For longer distances the convergence was better but thinking on an experiment we tried to keep the system as short as possible. 

Finally, the detection stage consist of a balanced photodetector, which was simulated measuring the intensity on each half of the grid.

\subsection{Post-processor training}\label{sec: PPT}

The method we used to train the programmable planes is known as \textit{wavefront matching}~\cite{sakamaki2007new}, which can be used to transform a given impinging wavefront onto a desired one at the output by choosing the spatially structured phase patterns on a set of programmable phase planes. The phases of the planes are modified in an iterative way, by propagating the impinging wavefront and the counter-propagating desired wavefront to each plane and setting the phases of the plane to match their difference at each pixel. As an example, this technique was used by J. Carpenter and coworkers to sort light depending on its spatial structure in the Laguerre(Hermite)-Gaussian mode basis~\cite{fontaine2019laguerre}. This technique was clearly illustrated by a set of tutorial videos~\cite{WFM_youtube}.

\section{Proof of concept: post-processor performance}

In this section, we quantify the resolving power of the pre- and post- processor with the proposed method. In particular, we give a clear example that shows how the inclusion of the pre-processor enhances the wavelength sensitivity of the whole device.

We start by considering only the post-processor and its performance in discriminating two wavelengths. In particular, we use the wavefront matching technique on three programmable planes to distinguish two different wavelengths for input beams with the same spatial structure. As target beams, we use a Gaussian beam centered at one side or the other of the balanced photodetector. 
Near 800~nm, we determined the minimum wavelength difference which the stand-alone post-processor can resolve is in the order of $\Delta\lambda=1$~nm. 
For wavelength differences above this value, the discrimination is almost perfect, more than 99\% of the light ends in the target mode. However, when reaching the limit, there is a considerable amount of light that impinges the \textit{wrong} detector. As an example of this, in  fig.~\ref{fig:Method}~(b) we show the case where both incident wavefronts at the first programmable plane are Gaussian beams with $\lambda$~=~$800$~nm and $\lambda$~=~$800.5$~nm. Here the wavelengths are only partially discriminated, as can be seen from the intensity 2D profile or the cut along the center of the image. The photodetectors show a large cross-correlation between the trained wavelengths showing that wavelengths separated by $\Delta \lambda$~=~$0.5$~nm are close to the limit of the post-processor.

Next we included the pre-processor to the propagation and training algorithm. In fig.~\ref{fig:Method}~(c) we show an example in which the light pass through the diffusive medium to generate a speckle pattern before the programmable planes. We use the same parameters as before, $\lambda=800$~nm and $\Delta\lambda=0.5$~nm.
Now we see that the two wavelengths can be discriminated with almost no spurious cross correlation. The percentage power that ends in the \textit{wrong} detector is below $0.1$~\%.  This example shown that the wavelength sensitivity of the post-processor is enhanced by the random pre-processing reducing the cross-correlation for two input light beams with wavelength differences below the limit the post-processor could accurately resolve.

As described, this method can optimally discriminate two wavelengths separated by approximately the coherence length $l_c$ of the pre-processor. If one would want to train the post processor in a range $\Delta \lambda^\mathit{tr}$ that is higher or lower than the coherence length, care must be taken to ensure proper wavelength discrimination, as we discuss in the next section, where we discuss how to use these results to construct a wide range wavemeter or, equivalently, a tunable wavelength division multiplexer.

\section{Wavemeter design: multi-range strategie}\label{sec: algorithms}

A wavemeter should be able, not only to discriminate between two specific wavelengths, but also to determine an unknown  wavelength over a given range. For the method described above, if the wavelength impinging the wavemeter is beyond the two trained target wavelenghts, the device will give an inconclusive result. To find a wavelength range the device could iterate between different post-processor configurations, i.e. trained wavelengths, until a positive measurement is obtained and the unknown wavelength is determined. To speed up this search it would be convenient to have the device working with variable precision. This way, to search for an unknown wavelength one would first use a wide precision and then narrow the measurement by increasing the resolution of the device.

In this section we now describe how to train the post-processor to determine wavelengths with different precision.
To extend the wavemeter optimal working precision range we study two training schemes. One where the post-processor is trained with only two wavelengths, and another one where we train it with multiple wavelengths, equally spaced between the extremes of the training range. To characterize both strategies we use speckle patterns with different correlation lengths and different training wavelength ranges. In the following subsections we show the results of the simulations for several parameter ranges, which allows us to determine the limits of both training methods and show how an optimal configuration should be chosen.

\subsection{Two wavelength scheme}

The first scheme consists in training the planes to recognize only the pair of wavelengths $\lambda_0$ and $\lambda_0 + \Delta \lambda^{tr}$, similar to what was described in the previous section.

\begin{figure}[ht]
    \centering
    \includegraphics[width=\columnwidth]{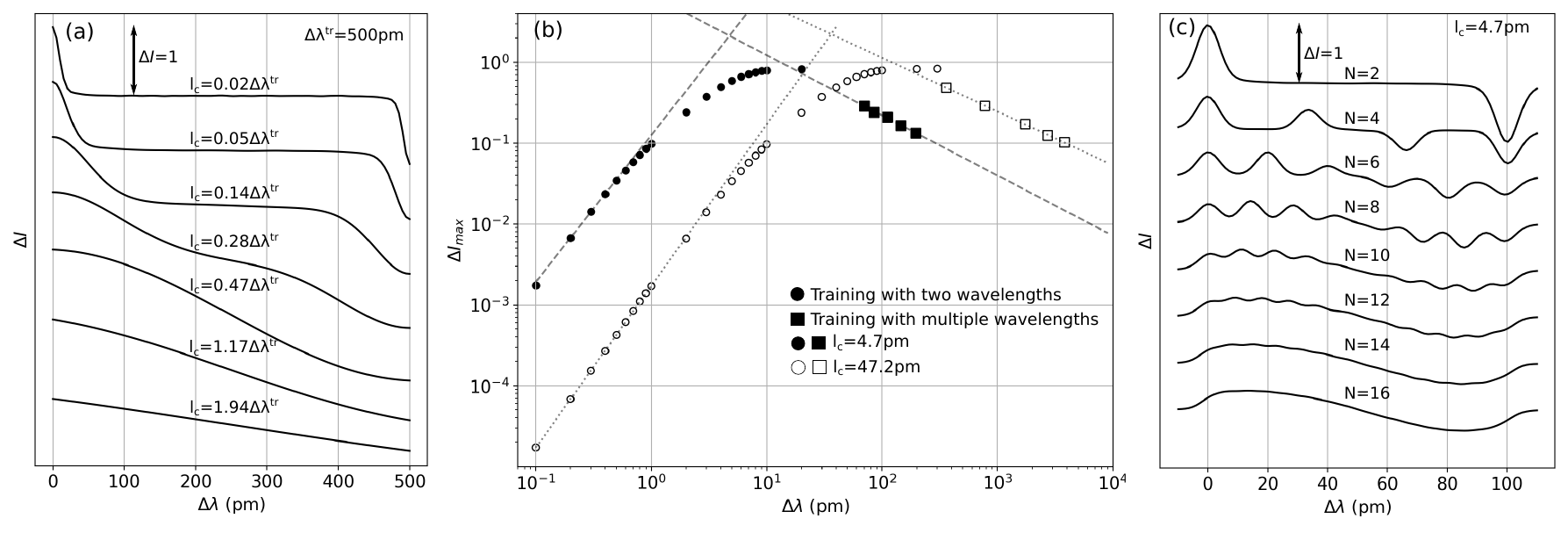}
    \caption{Simulation results. (a) typical results obtained in a simulation where only two wavelengths are used for training (in this case, 800~nm and 800.5~nm). $\Delta I = 1$ is shown to set a scale on the different curves as it shrinks for bigger correlation lengths. (b) characterization of the wavemeter where both training schemes are used to extend the dynamical range of the wavemeter. The circle points were obtained training only two wavelengths and the square points were obtained training enough wavelengths such that the curve between the trained wavelengths is smooth as can be seen in (c) for N~=~16. We characterized the system for two different correlation lengths at the output of the pre-processor and obtained similar performance of the wavemeter given by the slopes $\alpha$ of the pointed curves fitted to the asymptote of each curve. (c) typical results obtained in a simulation where more than two wavelengths are used in the training stage. In this case, for a total of N~=~16 trained wavelengths there is a monotone relationship between the wavelength and the normalized intensity difference measured by the detector.}
    \label{fig:Results}
\end{figure}

We define the normalized intensity difference between detectors as

\begin{equation}
    \Delta I = \frac{I_R - I_L}{I_R + I_L},
\end{equation}

where $I_R$ and $I_L$ are the total intensity in right and left detector. In fig.~\ref{fig:Results}(a) we show results for the intensity difference between detectors as a function of wavelength for a fixed training interval $\Delta \lambda^{tr}$~=~$500$~pm for the two wavelength training scheme. A qualitatively change in the behaviour of $\Delta I$ is observed along the interval as the correlation length $l_c$ of the generated speckle is varied. When $l_c \lesssim 2 \Delta \lambda^{tr}$ the information of $\Delta \lambda$ can be inferred from $\Delta I$ only in a region that is close to any of the trained wavelengths, in between there is a region of flat response, in which it will not be possible to associate a $\Delta I$ to a $\Delta \lambda$, as seen in the top 3 curves of fig.~\ref{fig:Results}~(a).

When $l_c$ grows, the flat region acquires a slope allowing to discriminate from this signal all the wavelengths in the range of $\Delta \lambda$, as seen in the middle curves of fig.~\ref{fig:Results}~(a). For bigger coherence lengths, the visibility of the signal is reduced, as seen in the bottom two curves. In this regime, when the coherence length is bigger than the training interval $l_c \gtrsim 2 \Delta \lambda^{tr}$, the two wavelength scheme will work correctly, with a varying sensitivity. 

To characterize the sensitivity of this scheme, in its working range, we record the difference of the normalized intensities on the detectors at the two extremes of the training range:
\begin{equation}
    \Delta I_{max} = I(\lambda_0)-I(\lambda_0+\Delta\lambda^\mathit{tr}).
\end{equation}

In fig.~\ref{fig:Results}~(b) we show the sensitivity $\Delta I_{max}$ as a function of the training range obtained in the simulations for $l_c$~=~$4.7$~pm and $l_c$~=~$47.2$~pm for $\lambda_0$~=~$800$~nm. The results for the two wavelength training are shown in circles. We fit the asymptotic behaviour with the function
$    \Delta I_{max}(\Delta \lambda) = (\Delta \lambda)^\alpha e^\beta     \label{eq:fit} $
where $\beta$ represents the intercept and $\alpha$ represents the slope. The parameter $\alpha$ is relevant for the characterization of the wavemeter because its value holds information on how pronounced the decay of $\Delta I_{max}$ is as $\Delta \lambda^{tr}$ gets narrower. The slopes $\alpha$ obtained from the fits for several $l_c$ were comparable, we report their average to be $\alpha_{2 \lambda} = 2.02\pm0.03$~fm$^{-1}$.

The two wavelength scheme described only works for $\Delta \lambda^{tr} \lesssim 2 l_c$, with decreasing sensitivity as $\Delta \lambda^{tr}$ becomes smaller that $l_c$. In this regime, electrical amplification of the signal can help increase resolution a few orders of magnitude, with the usual restriction on bandwidth and noise. On the other hand to extend the functionality of our wavemeter beyond this lower limit, into grater ranges where $\Delta \lambda^{tr} \gtrsim 2 l_c$, we discuss in the following section a multiple wavelength training method.

\subsection{Multiple wavelengths' scheme}

Here, we extend the training range of the post-processor by using not two but several wavelengths. The motivation for this proposal is to avoid the planar response observed for $N$~=~2 when $\Delta \lambda^{tr} \gtrsim 2 l_c$ to obtain a valid signal in this range. 

As an example in fig.~\ref{fig:Results}(c) we show $\Delta I(\Delta \lambda)$ obtained for a fixed training range $\Delta \lambda^{tr}$~=~$100$~pm and coherence length $l_c=4.7$~pm while varying the number N of trained wavelengths from 2 to 16. The training wavelengths are chosen spaced equally between $\lambda_0$ and $\lambda_0+\Delta \lambda^\mathit{tr}$. One can see that at a given number of trained wavelengths, in this case N~=~16, the response becomes monotone, not only eliminating the flat zone where the wavelength is undetermined, but also producing an ouput signal propotional to the wavelength difference over the specified range. The amount of training wavelenghts N required depends on the ratio  $\Delta \lambda^{tr}/l_c$, which can be roughly determined by the Nyquist–Shannon sampling theorem. As this factor grows, the number of wavelengths needed grows accordingly.

This way, we see this scheme allows us to train an interval in which $\Delta \lambda^{tr}\gg l_c$ and still have useful information in the whole range between trained wavelengths. Here also, as the training range increases, compared to the coherence length, the sensitivity of the signal decreases. This is shown in squares in fig.~\ref{fig:Results}~(b).
For this case, the asymptotic behaviour is considerably softer than for the two wavelength scheme and still roughly independent of the pre-processor coherence length, we now obtain  $\alpha_{N \lambda} = -0.61\pm0.06$~pm$^{-1}$.  We also note that in this case, the interval $\Delta \lambda$ in which one can associate unequivocally wavelength and intensity is slightly shorter than the trained interval $\Delta \lambda^{tr}$, this could be corrected by using a training region bigger than the target region.

\section{Discussion}

We presented two training schemes which work for different regions of $\Delta \lambda^{tr}$, a two wavelength scheme that works if $2 \Delta \lambda^{tr} \lesssim l_c $ and a multi wavelength scheme that works when $2 \Delta \lambda^{tr}\gtrsim l_c$. The dynamic range of a wavemeter constructed this way will be firstly determined by the measurement resolution achievable on measuring $\Delta I$. For a conservative sensibility of $10^{-3}$ the method works for more than two orders of magnitude above and below the coherence range. As described, one can extend this range by training different ranges, and determining by iteration  in which range the wavelength lies. 

Our simulations show that the decrease in sensitivity observed is a property of the optical post-processing stage and is independent of the correlation length of the speckle generated in the pre-processor. We also observe, that the dynamic range of the wavemeter is shifted with that correlation length $l_c$ but the asymptotic behaviour is roughly independent. In practice $l_c$ depends on the characteristics of the medium used as a pre-processor to generate speckle and can be taken as a constant.

\section{Conclusions}

We described how to construct a speckle wavemeter using a fixed scattering medium as a pre-processor and a programmable optical post-processor consisting of three programmable planes, which are trained with a wavefront matching technique. The device can be trained to determine a wavelength with a resolution way beyond the resolving power of the post-processor, and even above and beyond the coherence length of the first scattering medium. We propose a multi-range strategy and analyze two training methods: one that works better above the coherence length of the pre-processor and one that works better below. The experimental scheme that we propose replaces digital by optical processing therefore reducing significantly the measurement time typically needed for speckle based wavemeters. The proposed scheme has the potential to take the speckle based wavemeter-like setups to a next step in the active stabilization laser frequency, given that the wavelength can now be determined at a speed only determined by the bandwidth of a balanced photodetector and the feedback electronics.

\newpage

\begin{backmatter}
\bmsection{Funding}
This work was supported by Agencia I+D+i Grants No. PICT 2018 - 3350 and No. PICT 2019 - 4349, Secretaría de Ciencia y Técnica, Universidad de Buenos Aires Grant No. UBACyT 2018 (20020170100616BA), and CONICET (Argentina). 

\bmsection{Acknowledgments}
We would like to acknowledge Mikael Mazilu for pointing us into Ref. \cite{metzger2017harnessing} and sugesting one could replace digital processing for optical. We also thanks Nicolás Nuñez Barreto for careful reading of the manuscript.

\bmsection{Disclosures}
The authors declares no conflicts of interest.

\bmsection{Data availability} Data underlying the results presented in this paper are not publicly available at this time but may be obtained from the authors upon reasonable request.

\end{backmatter}


\bibliography{sample}






\end{document}